# Broadband terahertz generation via the interface inverse Rashba-Edelstein effect

C. Zhou[1,#], Y. P. Liu[2,3,#], Z. Wang[1], S. J. Ma[1], M. W. Jia[1], R.Q. Wu[1,4], L. Zhou[1], W. Zhang[5], M. K. Liu[6], Y. Z. Wu[1,*], J. Qi[2,*]

[1] Department of Physics, Fudan University, Shanghai 200433, China

[2] State Key Laboratory of Electronic Thin Films and Integrated Devices, University of Electronic Science and Technology of China, Chengdu 610054, China

[3] Institute of Modern Physics, Fudan University, Shanghai 200433, China

[4] Department of Physics and Astronomy, University of California, Irvine, CA 92697, USA

[5] Department of Physics, Oakland University, Rochester, Michigan 48309, USA

[6] Department of Physics, Stony Brook University, Stony Brook, New York 11794, USA

[#] These authors contributed equally to this work.

*e-mail: wuyizheng@fudan.edu.cn; jbqi@uestc.edu.cn

## Abstract

Novel mechanisms for electromagnetic wave emission in the terahertz (THz) frequency regime emerging at the nanometer scale have recently attracted intense attention for the purpose of searching next-generation broadband THz emitters. Here, we report a new mechanism for broadband THz emission, utilizing the interface inverse Rashba-Edelstein effect. By engineering the symmetry of the Ag/Bi Rashba interface, we demonstrate a controllable THz radiation (~0.1-5 THz) waveform emitted from metallic Fe/Ag/Bi heterostructures following photo-excitation. We further reveal that this type of THz radiation can be selectively superimposed on the emission discovered recently due to the inverse Spin Hall effect, yielding a unique film thickness dependent emission pattern. Our results thus offer new opportunities for versatile broadband THz radiation using the interface quantum effects.

Terahertz (THz) radiation from 0.1–30 THz accesses a diverse group of low-energy elementary excitations in solid-state systems [1], holding great promises for imaging, sensing and security applications [2]. One major challenge in the next generation THz technology is to search novel mechanism(s) providing efficient and broadband THz radiation with a gapless spectrum [3-5]. To date, most broadband THz emission devices [2-5] are based on the femtosecond laser excitations, taking advantage exclusively of the nonlinear or dynamic properties of the electrons. Recently, the emerging ultrafast spintronics [6-13], however, offers an alternative route to the THz emission with the spin-degree of freedom, by converting spin current bursts into THz pulses. In this way, one can effectively generate, control, and detect the spin currents, as well as utilize such spin-to-charge conversion [6-13] within the sub-picosecond timescale to yield efficient ultra-broadband THz emission. Such ultrafast spin-to-charge conversion process in all previous works is mostly based on the inverse Spin Hall effect (ISHE) (14-15), which happens inside the bulk of a metallic system with a strong spin-orbit coupling (SOC).

In contrast, the inverse Rashba-Edelstein effect (IREE) occurring at the interfaces with broken inversion symmetry can also provide efficient spin-to-charge conversion [16-19]. In the IREE, the generated charge current in two-dimensional electron gas can be described by [19]

$$\mathbf{j_c} \propto \lambda_{IREE} \mathbf{j_s} \times \hat{\mathbf{z}} \qquad (1)$$

where $\lambda_{IREE}$ is the IREE coefficient which is proportional to the Rashba parameter $\alpha_R$, $\hat{\mathbf{z}}$ is the direction of the potential gradient (interfacial electric field) perpendicular to the interface, and $\mathbf{j_s}$ is the spin current. Although the IREE has been intensively studied under equilibrium or quasi-equilibrium conditions in magnetoresistance measurements [17], non-local spin valves [18], and ferromagnetic resonance experiments [19], it is still elusive whether the IREE can work in femtosecond timescale, and play a vital role in the THz emission.

In this work, we report the observation of THz radiation via the interface IREE in the metallic Fe/Ag/Bi heterostructures, which clearly proves the effect of the interface IREE on the spin-to-charge conversion in femtosecond timescale. This observation brings us to a novel mechanism of emitting broadband THz wave from interface states, in contrast to the others based on the bulk properties dominating mainstream THz sources. In addition, the THz emission spectroscopy is also demonstrated to be an indirect way to reveal the evolution of the SOC-related electronic properties at the Ag/Bi interfaces.

Figure 1(a) shows the new scheme for an IREE-based THz emitter, which is a metallic heterostructure consisting of ferromagnetic (FM) and two non-ferromagnetic ($NM_1$ and $NM_2$) thin films. The Rashba SOC exists between $NM_1$ and $NM_2$. The FM film is magnetized by an in-plane magnetic field. A femtosecond laser pulse excites the heterostructure and leads to the generation of non-equilibrium spin-polarized electrons in the FM film [20], which subsequently super-diffuse across the neighboring $NM_1$, known as the spin current, and arrive at the $NM_1/NM_2$ Rashba interface. Due to the IREE, this longitudinal spin current $\mathbf{j}_s$ at the interface converts into a transient transverse charge current $\mathbf{j}_c$, which thereby acting as a source of THz radiation. Here, we select Ag and Bi as the two NM films, because the Ag/Bi interface is proved to have strong Rashba SOC between Bi(111) and Ag [19,21,22]. We first epitaxially grew a 2 nm Fe film on MgO(001) substrate, then grew either Ag/Bi or Bi/Ag bilayer on top of the Fe film. The *in-situ* reflection high-energy electron diffraction indicates that the Ag film on top of Fe film is single crystalline, but the Bi film is polycrystalline. The THz emission spectroscopy experiments were performed at room temperature. More experimental details about the samples and our THz emission spectroscopy setup are shown in the Supplemental Material [23].

Figure 1(b) illustrates the typical time-domain THz signals detected in two trilayer samples Fe(2)/Ag(2)/Bi(3) and Fe(2)/Bi(3)/Ag(2). The number inside each pair of brackets indicates the corresponding film thickness in nanometer. Obviously, the THz signals in these two samples are almost the same in shape and strength. A crucial observation is that the signals are completely out of phase once the stacking order of Bi/Ag is reversed. The THz signals from Fe/Bi and Fe/Ag bilayers are much smaller than those from Fe/Ag/Bi and Fe/Bi/Ag trilayer samples. Such observation is a direct evidence of THz radiation arising from the IREE-based spin-to-charge conversion. This is because when the stacking order of Bi and Ag is changed, the direction of the interfacial electric field is reversed (from $\hat{z}$ to $-\hat{z}$). According to Eq. (1), this will lead to a sign reversal of the charge current, and hence the THz emission, as demonstrated in Fig. 1(b). We note that this result is in sharp contrast to the signal reversal via ISHE [6, 9], where the direction of the spin current has to be reversed. In the current situation, the direction of the spin current is same since the relative position between the FM layer and the NM layers is not changed. It should be noted that, if rotating the in-plane magnetization, the angular-dependent THz signal shows a clear Sine-shape dependence (see Fig. 1(c)), which is expected by the IREE effect (see Eq.(1)).

Fig. 1(d) shows that a typical broad bandwidth from ~0.1 to 5 THz, peaking at around 2 THz, can be obtained in these samples. The maximum electric field in these two samples is evaluated to be ~1.3 V/cm, which is only about 6 times smaller than that in a ZnTe (110) crystal with a thickness of 1 mm measured under the same experimental conditions [23]. Such result is really remarkable if the THz signals from Fe/Ag/Bi are solely attributed to the IREE, because generally the effective thickness of the active Rashba interface is a monolayer (ML), which is almost 7 orders of magnitude smaller than that of the conventional nonlinear crystals. Strength of the peak electric field also suggests that IREE, in terms of the THz emission, is far superior to the other interface inverse spin-orbit torque effect [7] (Ref.7 used a laser fluence of 1 mJ/cm$^2$, which is larger than ours by two orders of magnitude.), and can be comparable with the bulk ISHE [6, 8-13]. Therefore, IREE may offer a new mechanism in generating the efficient broadband THz wave.

In order to further clarify and optimize the IREE-based THz emission, we have carried out control experiments on three samples: Fe(2)/Bi($d_{Bi}$), Fe(2)/Bi($d_{Bi}$)/Ag(2) and Fe(2)/Ag(2)/Bi($d_{Bi}$), where the Bi layer is wedge-shaped with its thickness $d_{Bi}$ increasing continuously from 0 to 4 nm. Figs. 2(a)-(c) illustrate the typical THz signals measured in these three samples with different thicknesses. In general, the Fe/Ag/Bi and Fe/Bi/Ag trilayers yield much stronger THz signals than Fe/Bi and Fe/Ag ($d_{Bi}$ or $d_{Ag}$ =0 nm, and also see Supplementary Material in [23]), which produce a comparable THz emission with a single layer of Fe ($d_{Bi}$=0 nm). These results can exclude the following effects dominantly contributing to the observed signals in trilayer samples: (1) demagnetization in Fe film [24], (2) ISHE in bulk Bi or Ag, (3) and potential signals from other interface states, e.g. Fe/Bi. The THz radiation via IREE at the Rashba Ag/Bi interface is further confirmed.

However, the THz signals obtained in Fe/Ag/Bi and Fe/Bi/Ag are not always out of phase, and even exhibit very different magnitudes in the low thickness regime. To understand such unusual behavior, we performed detail Bi-thickness dependent analysis of the trilayer samples, whose structures are simply depicted in Figs. 3(a)-(b). We define the amplitude $\Delta V$ as the peak-to-valley difference between the largest and smallest values in signal S(t). Under the current situation where the bandwidth, from ~0.1 to 5 THz, almost does not change with Bi thickness for each trilayer sample, the thickness dependence of THz signal can be largely manifested by $\Delta V$ as a function of $d_{Bi}$. $\Delta V$ ($d_{Bi}$) obtained in Fe/Ag/Bi and Fe/Bi/Ag are shown in Fig. 3(c). Dependence of $\Delta V$ on $d_{Bi}$ here is very different from that from the Co/Pt bilayers, where ISHE is the main mechanism for THz emission [23].

In an ideal IREE case, only the ultrafast spin current ($\mathbf{j_s}$) arrived at Rashba interface plays the role of THz emission (yellow regions in Fig. 3(a)-(b)). The spin current at Rashba interface through the nonmagnetic layer can be approximately proportional to $\exp(-d_{Ag}/\lambda_s^{Ag})$ for Fe/Ag/Bi or $\exp(-d_{Bi}/\lambda_s^{Bi})$ for Fe/Bi/Ag, where

$\lambda_s^{Ag}$ and $\lambda_s^{Bi}$ are the spin diffusion lengths in Ag and Bi, respectively. Simultaneously, the amplitude of the THz wave should also be strongly related to the optical pump in the FM Fe layer, which initiates the ultrafast spin current. This is confirmed by the results in Fig. 3(d), where the THz radiation is proportional to the pump laser fluence. This observation indicates that the photon-induced spin current density linearly depends on the optical intensity in the Fe layer, which, however, should be influenced by the thickness of NM layer as well. Consequently, we investigated such influence using the transfer matrix method [23] by considering the Fabry–Perot interference effect [8]. Because the hot-electron spin density in FM metals close to the FM/NM interface provides the dominant source of ultrafast spin current [6,20] (shaded regions in Figs. 3(a-b)), we can only focus on the optical intensity (or optical absorption) in the Fe part near Fe/Ag or Fe/Bi interface. We obtain that the optical intensity $I_O$ near the FM/NM interface almost decreases exponentially with NM layer thickness ($d_{NM}$) as $I_O(d_{NM}) \propto exp(-d_{NM}/\lambda_0^{NM})$, where $\lambda_0^{NM}$ is the decay constant. This equation implies that as the NM layer thickness increases, the spin current density reduces, due to the drop of laser intensity at the FM/NM interface. In Ref. 8, such optical intensity decay was also considered, which was assumed to be proportional to the optical absorption per unit thickness of the whole sample. In fact, our calculations show that when discussing the thickness-dependent THz signals, only the laser absorption matters while the absorption of the THz wave inside the samples can be neglected because its absorption length has a scale of ~100 nanometers [23].

Therefore, if Fe/Ag/Bi or Fe/Bi/Ag has the perfect IREE-based THz emission, $\Delta V$ as a function of $d_{Bi}$ can be approximately fitted by:

$$\Delta V \propto exp(-d_{Bi}/\lambda_{eff}), \qquad (2)$$

where $\lambda_{eff}$ is an effective decay constant. $\lambda_{eff}$ is defined as $\lambda_{eff} = \lambda_0^{NM}$ for Fe/Ag/Bi and $\lambda_{eff} = \left(\frac{1}{\lambda_0^{NM}} + \frac{1}{\lambda_s^{Bi}}\right)^{-1}$ for Fe/Bi/Ag, respectively. Here, $\lambda_0^{NM}$ characterizes the laser intensity drop due to the optical absorption with the total thickness $d_{NM}$ (= $d_{Bi} + d_{Ag}$) of NM layers increasing. The fitted results are illustrated in Fig. 3(c). Clearly, Eq. (2) can well reproduce $\Delta V$ for Fe/Ag/Bi at $d_{Bi} > 0.5$ nm and for Fe/Bi/Ag at $d_{Bi} > 3$ nm, respectively, which can be used to define regimes with a well-established Rashba interface. According to the fittings, $\lambda_{eff}$ is obtained to be ~15.6 nm for Fe/Ag/Bi and ~7.8 nm for Fe/Bi/Ag. If $\lambda_0^{NM}$ is assumed to be the same for both samples, the spin diffusion constant $\lambda_s^{Bi}$ can be derived to be ~15.6 nm. However, for $d_{Bi} < 0.5$ nm in Fe/Ag/Bi and $d_{Bi} < 3$ nm in Fe/Bi/Ag, the measured $\Delta V(d_{Bi})$ data show considerable deviation between the exponential fitting and the experiments.

In Fig. 3(c), a sharp increase and decrease of $\Delta V$ across ~0.3 nm is observed in both trilayer samples. As stated, Eq. (2) is valid only if a perfect Rashba interface exists between Ag and Bi at all $d_{Bi}$. In reality, such condition can be hardly satisfied. Particularly, a prominent phenomenon discovered previously shows that upon deposition of the first few monolayers Bi, there could appear the AgBi surface alloying with strong SOC [21,22,25]. This is similar to the diluted CuBi alloy system, where a small amount of Bi impurities (<0.5%) in Cu could induce a large spin Hall effect [26]. Therefore, we also expect the existence of non-negligible SHE when $d_{Bi}$ is very small. In order to further understand the spin Hall conductivity (SHC) of Bi/Ag bilayers with different Bi thickness, we conducted the first-principle density functional theory calculations. Our calculations were performed using the projector augmented wave method [27,28] as implemented in the Vienna ab initio simulation package (VASP) [29,30]. The electronic exchange-correlation were described within a spin-polarized generalized-gradient approximation [31] with an energy cutoff of 350 eV. The SHC was evaluated by the Kubo formula [32], and a fine 36×36×1 k-point mesh in the full Brillouin zone and 720 bands were adopted to ensure the numerical convergence. The Ag(001) substrate was simulated by a 6-layer slab with a 3×3×1 supercell. We placed different numbers of Bi atoms on the Ag surface (see Fig. 4(a)-(b)), which can be regarded as different Bi coverage in the experiments. Typically, we performed the calculation with the 1/9, 4/9, 1, and 2 ML Bi coverage. The calculated intrinsic SHC in Fig. 4(c) demonstrates that the SHC reaches its maximum (1931 $\Omega^{-1}cm^{-1}$) at 1/9 ML Bi, and is greatly suppressed by further Bi coverage. This result is consistent with the observation that the THz radiation experiences a sharp increase and decrease across a sub-monolayer Bi thickness, i.e. ~0.3 nm. We also decomposed the SHC contribution from each individual element by selectively switching off the spin-orbital interaction of Bi and Ag atoms. Fig. 4(c) indicates that the SHC is mainly contributed by Bi atoms for Bi deposition of less than 1 ML. Although during the calculation we only considered the case where the Bi atoms grow on top of the Ag film, the remarkable SHC can also be expected to exist in the system with Ag atoms grown on top of Bi.

The Rashba effect in Bi/Ag(001) system can also be revealed by the theoretical calculation. Fig. 4(d) shows that the calculated spin-projected band structures along –X→Γ→X of 1 ML Bi on Ag(001) surface. Due to the hybridization between the Ag s-electron and the Bi p-electron, there is a clear spin-orbit splitting in the bands across a large energy range. This band splitting only locates inside the Bi layer and the first Ag layer, confirming the existence of the Rashba effect due to the inversion-symmetry-broken at the Bi/Ag interface with thickness-scale of 1 ML. Clearly, the THz radiation via IREE will involve a 180-degree phase change by altering the Bi/Ag interface symmetry via switching the Bi and Ag stacking order.

Therefore, for 0< $d_{Bi}$ <0.5 nm we have observed a superposition of the THz signals via ISHE and IREE in both samples. Such superposition can be manipulated by controlling the symmetry of Ag/Bi interface. In specific, the superposition is constructive and destructive for Fe/Ag/Bi and Fe/Bi/Ag, respectively. This finding well explains the sign change of ΔV in Fe/Bi/Ag. However, in this sample, quantitatively describing the gradual thickness-dependent signal for $0.5 < d_{Bi} < 3$ nm is still quite challenging. A qualitative description is given as follows. In our experiments, the Fe film is grown epitaxially on MgO(001) substrate. The subsequent growth of Ag on Fe(001) is also epitaxial. In contrast, due to the large lattice mismatch between Bi and Fe, Bi grown on Fe(001) film is mostly polycrystalline. This leads to the subsequent non-epitaxial growth of Ag film. Then, the Bi/Ag Rashba interface for Fe/Bi/Ag is expected to slowly establish as the number of the deposited Bi atoms increases, in contrast to the case of Fe/Ag/Bi. Thus, the corresponding critical Bi thicknesses to fully develop the Rashba interface are ~3 nm and ~0.5 nm for Fe/Bi/Ag and Fe/Ag/Bi, respectively. Therefore, accurate modeling of the Rashba interface requires further investigations using different experimental and theoretical approaches. Nevertheless, our findings provide indirect evidence of the gradual evolution of electronic properties associated with SOC at Ag/Bi interfaces.

In summary, our results clearly prove that the IREE can induce the ultrafast spin-to-charge conversion at sub-nanometer interfaces, which can also provide a new effective way for generating broadband THz emission at room temperature. The bandwidth of such THz wave dominantly covers ~0.1 to 5 THz. Its phase can be changed by controlling the inversion symmetry at the Ag/Bi and Bi/Ag interfaces. Thus, our work provides an encouraging example of how the THz emitters via interface Rashba effect can achieve performances close to their bulk counterparts. If properly engineering the heterostructure with larger spin-to-charge conversion effect at interfaces, much stronger THz wave may be expected. One possible candidate might be the heterostructures composed of the ferromagnetic film and the topological insulators because the metallic surface states of the topological insulators possess an intrinsic property of the spin-momentum locking [33,34], which could amplify the ultrafast spin-to-charge conversion efficiency. Our experiments also demonstrate that THz radiation arising from both IREE and ISHE can superimpose with each other.

Therefore, the emission efficiency can be further improved by utilizing both effects in the same device. Finally, our results demonstrate the THz emission spectroscopy to be a useful way to reveal the evolution of the electronic properties associated with the SOC at interfaces or surfaces, where many peculiar quantum effects often emerge.

**Acknowledgments**

This research is supported by the National Key Basic Research Program of China (Grant No. 2015CB921401), the National Key Research and Development Program of China (Grant No. 2016YFA0300703, 2017YFA0303504), the National Science Foundation of China (Grants No. 11734006, No. 11474066, No. 11434003, No. 11674068), Science Challenge Program of China (TZ2016004), and the Program of Shanghai Academic Research Leader (No. 17XD1400400).

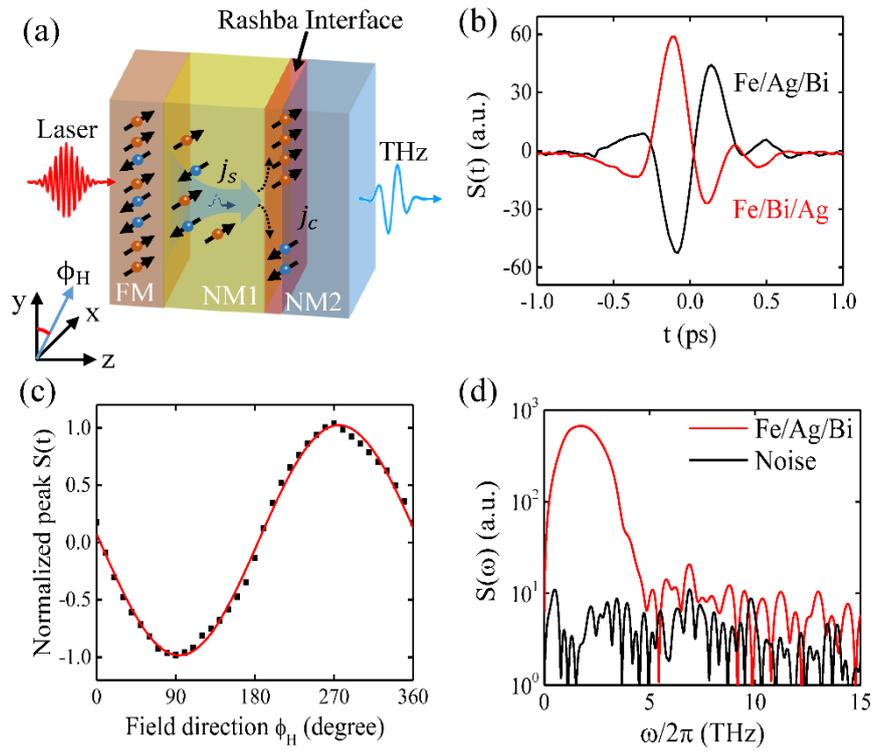

Fig. 1 (a) Schematic of THz emission via IREE upon excitation of ultrafast laser pulses. (b) Typical time-domain THz signals in Fe(2)/Ag(2)/Bi(3) and Fe(2)/Bi(3)/Ag(2) films. (c) Normalized peak S(t) versus direction of the external magnetic field together with a sinusoidal curve, measured on a Fe(2)/Ag(2)/Bi(3) film. As described in (a), the field **H** is applied within the sample plane (x-y). Its direction is defined by the angle $\phi_H$ between **H** and +y. THz detection is optimized along y-axis. (d) The frequency-domain THz signal in Fe/Ag/Bi corresponding to the time-domain signal in (b). Background noise is also shown in (d). The number inside each pair of brackets indicates the corresponding film thickness in nanometer.

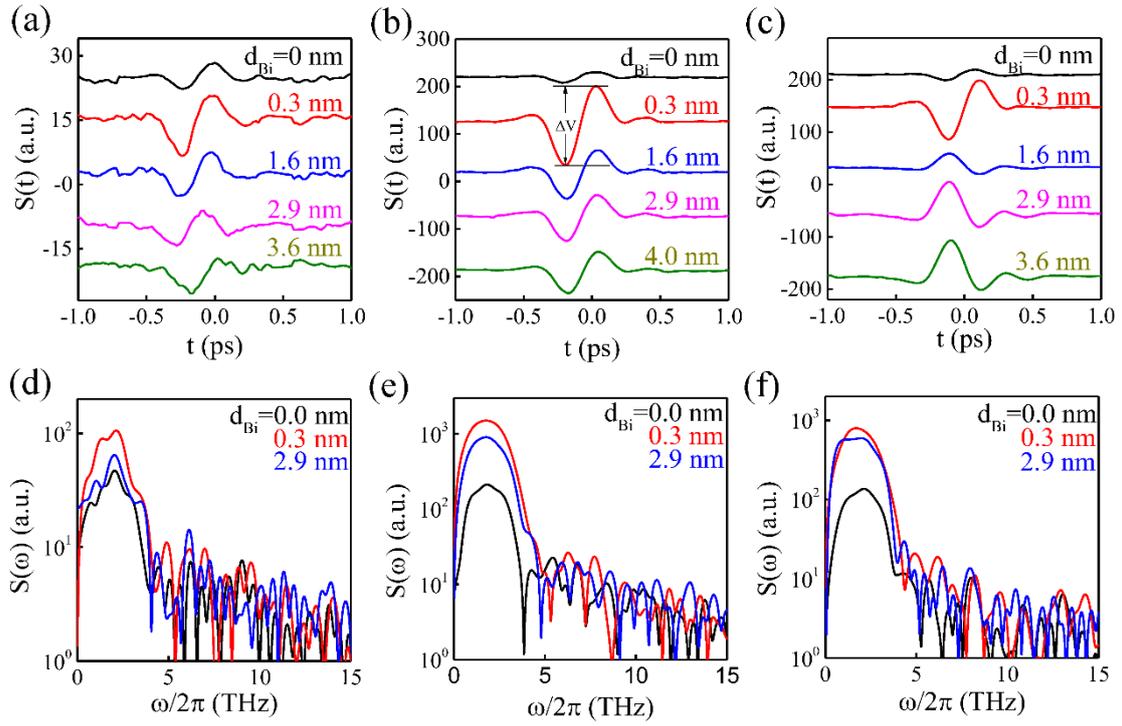

Fig. 2 Typical time-domain THz signals at different Bi thickness for (a) Fe(2)/Bi(wedge), (b) Fe(2)/Ag(2)/Bi(wedge), and (c) Fe(2)/Bi(wedge)/Ag(2). The corresponding frequency-domain signals are shown in (d-f), respectively. The number inside each pair of brackets indicates the corresponding film thickness in nanometer.

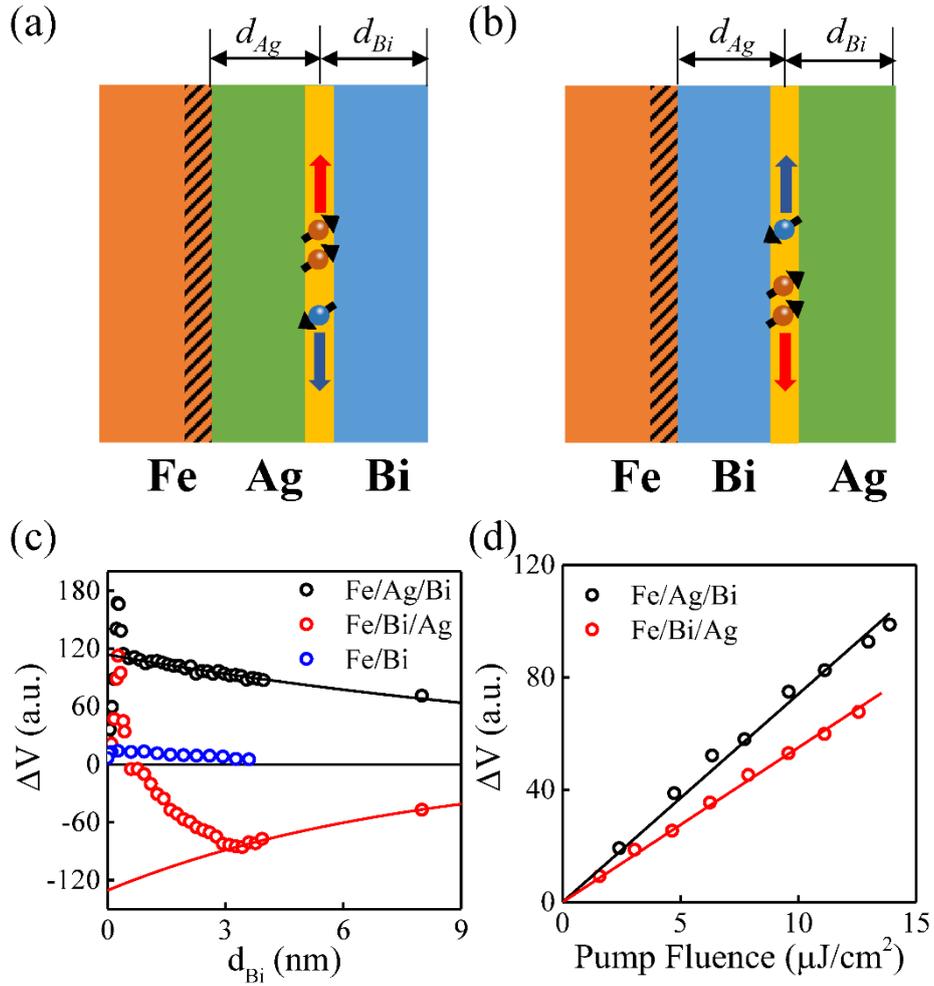

Fig. 3 Schematics of key parameters determining the thickness dependence of THz signals in (a) Fe/Ag/Bi and (b) Fe/Bi/Ag. The shaded region in FM layer represents the portion near the FM/NM interface dominating the spin current injection into the NM layers. The yellow regions indicate the Rashba interfaces. Switching the stacking sequence of Bi and Ag films induces the direction reversal of the net transverse charge current at the Rashba interface, as indicated by the red and blue arrows. (c) Peak-to-peak amplitude ΔV as a function of Bi thickness in Fe/Ag/Bi, Fe/Bi, and Fe/Bi/Ag films. The solid lines are fitted curves using Eq. 2. (d) Peak-to-peak amplitude ΔV as a function of laser pump fluence in Fe(2)/Ag(2)/Bi(2) and Fe(2)/Bi(2)/Ag(2). Solid lines are linear fittings.

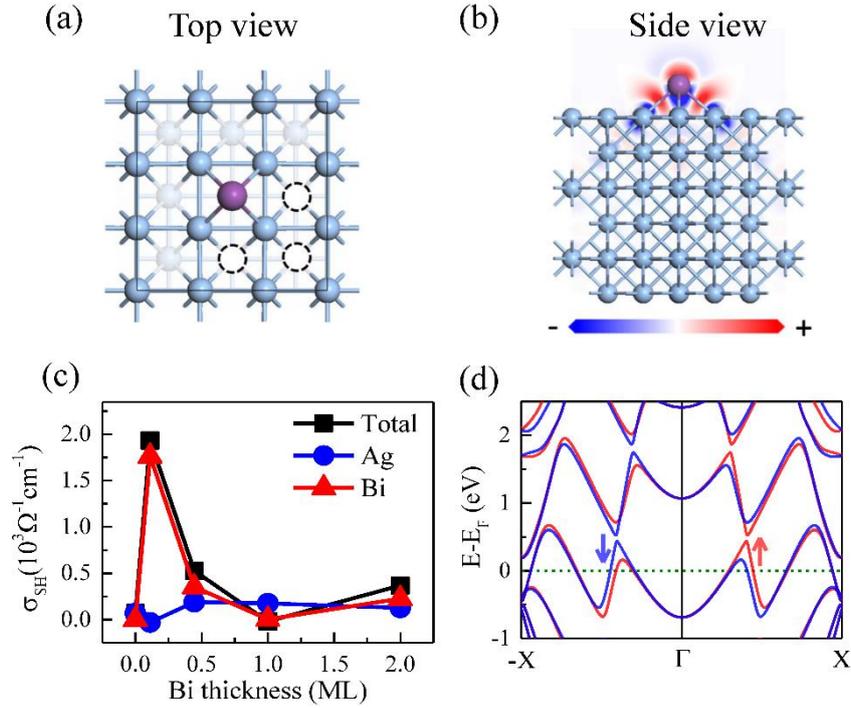

Fig. 4 (a) and (b) Schematic of the supercell for calculations of Bi atoms deposited on Ag(001) from the top view and side view, respectively. The color inside (b) represents the charge redistribution between the adsorbed Bi atom and Ag(001) surface. (c) Spin Hall conductivity as a function of Bi thickness associated with Ag, Bi and total (Ag+Bi), respectively. (d) The spin-resolved electronic band structure of Bi(1 ML)/Ag(001) along Γ-X direction, with the red (blue) color representing the spin-resolved band with the spin polarization along +y (-y) direction.